\documentclass[aps,prl,twocolumn,superscriptaddress]{revtex4-2}
\usepackage[utf8]{inputenc}
\usepackage[T1]{fontenc}
\usepackage[english]{babel}
\usepackage{graphicx}    % For figures
\usepackage{dcolumn}     % Align table columns on decimal point
\usepackage{bm}          % Bold math symbols
\usepackage{amsmath, amssymb, amsfonts}
\usepackage{textcomp,amsmath,amssymb}    % Special symbols
\usepackage{float}
\usepackage{microtype}
\usepackage[titletoc]{appendix}
\usepackage[section]{algorithm}
\usepackage{emptypage}
\usepackage[table]{xcolor}
\usepackage{csquotes}
\usepackage{stackengine} 
\usepackage{subfigure}
\usepackage{siunitx}
\usepackage{multirow}
\usepackage[pdfa]{hyperref}
\usepackage[dvipsnames]{xcolor} 
\begin{document}
\title{Is the large uncertainty of $\delta_{CP}$ fundamentally encoded in the neutrino quantum state?}
% Title and authors
%\title{Unveiling the Precision Boundaries in Neutrino Physics: \\ A Quantum Metrology Perspective}
\author{Michela Ignoti}
\email{michela.ignoti@studenti.unimi.it}
\affiliation{Dipartimento di Fisica “Aldo Pontremoli”, Università degli Studi di Milano, I-20133 Milano, Italia}
\affiliation{INFN Sezione di Milano, I-20133 Milano, Italia}
\author{Claudia Frugiuele}
\email{claudia.frugiuele@mi.infn.it}
\affiliation{INFN Sezione di Milano, I-20133 Milano, Italia}
\author{Matteo G. A. Paris}
\email{matteo.paris@fisica.unimi.it}
\affiliation{Dipartimento di Fisica “Aldo Pontremoli”, Università degli Studi di Milano, I-20133 Milano, Italia}
\affiliation{INFN Sezione di Milano, I-20133 Milano, Italia}
\author{Marco G. Genoni }
\email{marco.genoni@unimi.it}
\affiliation{Dipartimento di Fisica “Aldo Pontremoli”, Università degli Studi di Milano, I-20133 Milano, Italia}
\affiliation{INFN Sezione di Milano, I-20133 Milano, Italia}
%%%
\date{\today}
% Abstract
\begin{abstract}
The precise measurement of the leptonic CP-violating phase $\delta_{CP}$ remains one of the major open challenges in neutrino physics, as current experiments achieve only very limited accuracy. We address this issue through the lens of quantum estimation theory. A distinctive feature of neutrino oscillation experiments is that they cannot freely optimize the probe or measurement, since both are constrained by the production and detection of flavor eigenstates. We therefore examine whether the large uncertainty in $\delta_{CP}$ originates from intrinsic  reasons, either of the neutrino quantum state or of flavor measurements, or if it instead stems from experimental limitations. By comparing quantum and classical Fisher information, we demonstrate that the limited sensitivity to $\delta_{CP}$ originates primarily from the information content of flavor measurements. Furthermore, we  show that targeting the second oscillation maximum, as in the ESS$\nu$SB proposal, substantially enhances $\delta_{CP}$ information compared to experiments centered on the first maximum.
%The precise measurement of the leptonic CP-violating phase $\delta_{CP}$ remains one of the major open challenges in neutrino physics. We apply quantum estimation theory to determine whether this uncertainty is intrinsic to the neutrino quantum state. Computing the Quantum Fisher Information (QFI), we find that the state itself contains sufficient information, while the flavor Fisher Information (FI) shows that current strategies at the first oscillation peak are highly suboptimal. Crucially, the FI grows at the second peak, suggesting that next-generation experiments can dramatically improve precision. In particular, the proposed ESS$\nu$B facility is uniquely positioned to make a decisive breakthrough in the determination of $\delta_{CP}$.
\end{abstract}
\maketitle
\textit{\textbf{Introduction.—}}
Neutrinos are among the most elusive particles in nature, interacting only via the weak force. Although the Standard Model predicts them to be massless, the observation of neutrino oscillations demonstrates that they possess nonzero masses. These oscillations are described by the Pontecorvo–Maki–Nakagawa–Sakata (PMNS) matrix, parameterized by three mixing angles ($\theta_{12}$, $\theta_{13}$, $\theta_{23}$) and a Dirac CP-violating phase $\delta_{\mathrm{CP}}$. While the mixing angles and mass-squared differences ($\Delta m^2_{21}$, $\Delta m^2_{31}$) are now determined with percent-level precision, $\delta_{\mathrm{CP}}$ remains poorly constrained. The NuFIT~v6.0 global analysis reports \cite{Esteban2024NuFit}, for the normal ordering (NO) scenario, $\delta_{CP}^{(\mathrm{NO})} = 177^{\circ}\,^{+19^{\circ}}_{-20^{\circ}}$, based primarily on data from the long-baseline accelerator experiments T2K~\cite{Abe_2023} and NO$\nu$A~\cite{PhysRevD.106.032004}, which produce intense neutrino beams from proton collisions and measure their flavor composition after propagation over hundreds of kilometers. The precision on $\delta_{CP}$ remains substantially lower than that of the other mixing parameters, for example, the mixing angle $\theta_{13} = 8.52^{\circ}\,^{+0.11^{\circ}}_{-0.11^{\circ}}$, highlighting the current experimental challenges in probing CP violation in the neutrino sector. Next-generation facilities such as DUNE~\cite{dune, universe10050221}, T2HK~\cite{yokoyama2017hyperkamiokandeexperiment, singh2023flavordependentlongrangeneutrinointeractions,universe10050221} and ESS$\nu$SB~\cite{Alekou_2023, Alekou2021}  aim to significantly improve this precision and may provide the first definitive evidence of leptonic CP violation.
A nonzero Dirac CP-violating phase $\delta_{\mathrm{CP}}$ in the PMNS matrix would signal CP violation in the lepton sector, a key ingredient for leptogenesis \cite{leptoG}, and thus supports scenarios connecting low-energy neutrino parameters to high-energy dynamics. Consequently, understanding the origin of the present uncertainty is not merely a technical matter but one with broad implications for particle physics and cosmology.

Quantum estimation theory (QET)~\cite{paris2009,Helstrom:1969fri} provides a rigorous framework in order to approach this challenge both from a fundamental and practical level. Let us given a quantum state $|\psi_\lambda\rangle$ that parametrically depends on a parameter $\lambda$ that we aim to estimate, and let us suppose to perform a quantum measurement described by a positive-operator valued measure (POVM) $\{\Pi_x\}$, the variance of any unbiased estimator of the parameter $\tilde{\lambda}$ is lower bounded by the Cram\'er-Rao bound (CRB)
\begin{equation}
    \mathrm{Var}_{\tilde{\lambda}}(\lambda) \geq \frac{1}{M F(\lambda)} \,.
    \label{eq:crb}
\end{equation}
Here $M$ corresponds to the number of independent measurements, while
\begin{equation}
F(\lambda) = \sum_x \frac{(\partial_\lambda p(x|\lambda))^2}{p(x|\lambda)} \,,
\label{eq:fi_app_def}
\end{equation}
denotes the (classical) Fisher information (FI), written in terms of the conditional probability $p(x|\lambda)=\langle\psi_\lambda | \Pi_x |\psi_\lambda\rangle$. The FI clearly depends on both the state
$|\psi_{\lambda}\rangle$ and on the POVM operators $\{\Pi_x\}$ describing the measurement strategy. Remarkably one can optimize over all the possible POVM operators, and obtain a measurement-independent quantity. In turn, the quantum Fisher information (QFI), defined as $H(\lambda) = \max_{\{\Pi_x\}} F(\lambda)$, is given by 
\begin{align}
H(\lambda) = 4 \left[ \langle 
\partial_{\lambda}\psi_\lambda | 
\partial_{\lambda}\psi_\lambda \rangle  - | 
\langle \partial_{\lambda}\psi_\lambda |\, \psi_\lambda \rangle 
|^2\right] \,.
\label{eq:qfi_generic}
\end{align}
As it is apparent from its definition, the QFI sets a more fundamental bound, the so-called quantum Cram\'er-Rao bound (QCRB)
\begin{equation}
    \mathrm{Var}_{\tilde{\lambda}}(\lambda) \geq \frac{1}{M F(\lambda)} \geq \frac{1}{M H(\lambda)}  \,,
    \label{eq:qcrb}
\end{equation}
which in the final instance depends only on the quantum state $|\psi_\lambda \rangle $, and not on the particular measurement strategy implemented.
Comparisons between the FI and QFI are
therefore instructive: if they coincide, the precision limit
is intrinsic; if a significant gap exists, the measurement
strategy is suboptimal. 
Standard QET applications focus on designing optimal probes and measurements that optimize and saturate the quantum Cramér–Rao bound, thereby achieving the ultimate precision allowed by quantum mechanics. These approaches have been instrumental in order to conceive and assess quantum sensing protocols promising a quantum enhancement in the estimation of physical parameters~\cite{giovannettiNatPhoton,RMPQuantumSensing,PezzeQuantumSensing,PerspectiveMultiPar,ReviewVictor}. Tools from QET have been employed in high energy physics, for example in the context of the anti-de Sitter/conformal field theory (AdS/CFT) correspondence~\cite{bak2016information,lashkari2016canonical,trivella2017holographic,QETcosmology2018,QETcosmology2020,erdmenger2020information,QETcosmology2021}, or to assess an Unruh-DeWitt detector~\cite{feng2022quantum}, universal gravity corrections \cite{candeUGC}, equivalence principle \cite{seveWEP}, and gravitational time dilation \cite{cepoTD}.
More recently, they have been exploited 
to derive and discuss the ultimate limits on the precision of physical parameters describing a QED 
scattering experiment~\cite{AlessioQED} and cosmological parameters~\cite{AlessioCosmological}.

This Letter, together with a upcoming paper applying QET to all relevant neutrino oscillation parameters~\cite{upcoming}, aims to provide a quantitative framework for assessing precision limits in the estimation of the CP-violating phase $\delta_{CP}$, and guiding the design of future experiments. 
%\marco{Earlier attempts in this direction were pursued in a very simplified scenario~\cite{nogueira2016quantumestimationneutrinooscillations}, i.e. by assuming neutrinos described by a two-dimensional Hilbert space, discussing the ultimate limits on the estimation precision for the mixing angle between the two neutrino's flavors, and clearly without any reference to the estimation of the $\delta_{CP}$ phase.}
Unlike typical quantum metrology protocols, neutrino oscillation experiments operate in a fundamentally different regime: neither the probe nor the measurement can be optimized, as both are fixed by the flavor production and detection processes. This constraint implies that intrinsic precision limits may arise from both the quantum state and the measurement. In this context, the central question is not how to engineer optimal settings, but what precision is fundamentally attainable under unavoidable constraints in the design of the estimation protocol. In particular, we investigate whether the restriction to flavor measurements imposes an additional limitation, beyond the intrinsic quantum bound, on the estimation of $\delta_{CP}$. 
%This Letter, together with a upcoming paper applying quantum estimation theory to all relevant neutrino oscillation parameters~\cite{upcoming}, provides a quantitative framework for assessing precision limits and guiding the design of future experiments.

Since we address the CP-violating phase $\delta_{CP}$, we focus our attention to accelerator-based experiments. To evaluate the impact of the restriction to flavor states and flavor measurements, we work under a set of simplifying assumptions. We consider neutrinos propagating in vacuum and model their quantum states as pure plane-wave states~\cite{Akhmedov:2009rb, Eliezer:1975ja, Bilenky:1976yj, Fritzsch:1975rz}. 
This restriction further limits our study to experiments where matter effects are negligible, such as T2K, T2HK~\cite{T2K:2011qtm, Abe_2023}, and the proposed ESS$\nu$SB~\cite{Alekou2021, Alekou_2023}, and we also assume a monochromatic neutrino beam at its peak energy $E$, which constitutes a reasonable approximation for these setups.

Under these assumptions, a neutrino produced at time $t_0=0$ in a flavor eigenstate $|\nu_\alpha\rangle$ evolves into $|\nu_\alpha(t)\rangle$ at time $t$, yielding the transition probabilities $P(|\nu_\alpha\rangle \!\to\! |\nu_\beta\rangle)=|\langle \nu_\beta | \nu_\alpha(t)\rangle|^2$. 
The corresponding expressions for antineutrinos, $|\bar\nu_\alpha(t)\rangle$ and $P(|\bar\nu_\alpha\rangle \!\to\! |\bar\nu_\beta\rangle)$, are obtained analogously. 
Further details on these, which form the basis of our analysis, are provided in the end matter. In general, these quantities will depend on the parameters entering in the PMNS matrix, the squared mass differences $\Delta m_{ji}^2 = m_j^2 - m_i^2$ of mass eigenstates, and on the ratio $L/E$ between the accelerator baseline $L$ and the beam peak energy $E$. As we are interested in the precision bounds on $\delta_{CP}$, we will consider all the other parameters fixed according to the NuFit best values~\cite{Esteban2024NuFit} (as regards the mass differences, we focus on the normal mass ordering (NO), but our results are practically identical for the inverted ordering (IO)). In particular we will then consider the behavior of the corresponding QFIs and FIs as a function of the ratio $L/E$, as it allows to properly identify the performances of the different current and future accelerator experiments designed to estimate $\delta_{CP}$. 
\\
%
%Introducing more realistic elements, such as matter effects, multiparameter correlations, decoherence, and detector inefficiencies, would reduce both the QFI and the FI; therefore, the bounds presented here should be interpreted as lower limits on the achievable precision. 
%MAGARI NN DIMINUISCONO ALLO STESSO PASSO E QUINDI LA FI DIVENTA QUASI OTTIMALE??

\textit{\textbf{Evaluating the QCRB —}}
The CP-violating phase is not directly measurable and its value is inferred from flavor-dependent detection rates determined by the oscillation pattern of neutrino propagation. We here proceed to derive the fundamental limits on its estimation precision by evaluating the QFIs corresponding to experiments based on both initial muon-neutrino and muon-antineutrino beams. This can be directly evaluated via Eq.~\eqref{eq:qfi_generic} by considering  $\lambda=\delta_{CP}$ as the parameter we want to estimate, and by replacing $|\psi_\lambda\rangle$ with either $|\nu_\mu(t)\rangle$ and $|\bar\nu_\mu(t)\rangle$. We denote respectively with $H_\nu(\delta_{CP})$ and $H_{\bar\nu}(\delta_{CP})$ the corresponding QFIs. 
As neutrino and antineutrino experiments are statistically independent, the ultimate bound on the estimator variance ${\rm Var}_{\tilde{\delta}_{CP}}(\delta_{CP})$ in Eq.~\eqref{eq:qcrb}, taking into account both beams, is obtained by replacing the denominator with:
\begin{align}
M H_{\nu,\overline{\nu}}(\delta_{CP}) = M (\xi_\nu H_{\nu}(\delta_{CP}) + \xi_{\bar\nu} H_{\bar\nu}(\delta_{CP})) \,,
\end{align}
where $\xi_\nu = M_\nu / M$ and $\xi_{\bar\nu} = M_{\bar\nu}/M$ are respectively the fractions of the total number of measurements $M=M_\nu + M_{\bar\nu}$ corresponding to either neutrino ($M_{\nu}$) or antineutrino ($M_{\bar\nu}$) events. Our calculations show that, in the parameter regime of interest, $H_{\nu}(\delta_{CP})$ is approximately equal to $H_{\bar\nu}(\delta_{CP})$. Consequently, we can safely quantify the information contained both in neutrino and antineutrino quantum states simply as the average of the individual QFIs: $H_{\nu,\overline{\nu}}(\delta_{CP}) = (H_{\nu}(\delta_{CP})+H_{\bar\nu}(\delta_{CP}))/2$. We remark that the total number of events $M$ corresponds to the so-called {\it non oscillated} (anti)neutrinos, that is the total number of events in absence of oscillation phenomena, formally obtained by assuming that the PMNS matrix is the identity, and experimentally efficiently estimated from the number of events obtained at the near-detector.

\begin{figure}[H] %h per here se no la mette dove vuole lui
 
    \centering
    \subfigure{
    \includegraphics[width=0.45
\textwidth, height=0.6\textheight, keepaspectratio]{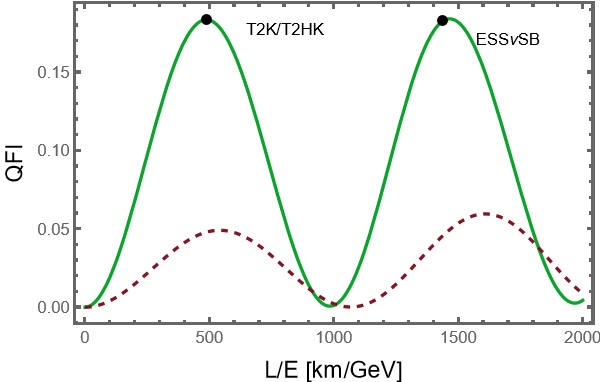}}
    
    \caption{The effective QFI $H_{\nu,\overline{\nu}}(\delta_{CP})$  (green solid line) and the neutrino appearance probabilities $P(\nu_\mu \to \nu_e)$ (dashed bordeaux line) plotted as a function of $L/E$ and for $\delta_{CP}=\delta_{CP}^{(\mathrm{NO})}$.  The two black points correspond to the value of the QFI for the  experiments T2K/T2HK and ESS$\nu$SB, identified by considering the peak energy $E$ of the neutrino spectrum.}
   \label{fig:qfi-delta}
\end{figure}

We find that $H_{\nu,\overline{\nu}}(\delta_{\mathrm{CP}})$ exhibits no significant dependence on $\delta_{\mathrm{CP}}$, indicating that the information content of the quantum state is largely insensitive to the true value of the parameter. Fig.~\ref{fig:qfi-delta} shows the QFI for $\delta_{\mathrm{CP}}$, evaluated at the NuFIT best-fit value $\delta_{CP}^{(\mathrm{NO})}$: we observe that the QFI peaks remain stable under variations in $L/E$, demonstrating consistent sensitivity across oscillation maxima. These peaks coincide with the oscillation maxima of \( P(\nu_\mu (\overline{\nu}_\mu) \to \nu_e (\overline{\nu}_e)) \), as the highest appearance probability provides maximal information on \( \delta_{CP} \). Both T2K and ESS$\nu$SB operate near the QFI maximum, making them equally well suited for \( \delta_{CP} \) measurement from the standpoint of the quantum state's information content. From the corresponding QCRB, we obtain a fundamental limit on the achievable precision of the CP-violating phase: for a total data sample of $M=\mathcal{O}(10^3)$ neutrino and antineutrino events, representing a conservative estimate of current statistics, the corresponding QFI yields a lower bound of approximately $3^\circ$ on the uncertainty of $\delta_{CP}$. If we take the current best-fit value $\delta_{CP}^{(\mathrm{NO})}$, the resulting uncertainty is at the few-percent level, indicating that the neutrino quantum state intrinsically encodes sufficient information to allow precision measurements even with present-day event statistics.

%\\[0.7\baselineskip]
\textit{\textbf{Evaluating the CRB for flavor measurement —}}\label{Fisher Information}
While the QFI sets the ultimate quantum limit, it is crucial to assess the classical FI Eq.\eqref{eq:fi_app_def} associated with flavor measurement strategies. To express this in the language of QET, we consider a positive operator-valued measure (POVM) corresponding to projections onto flavor eigenstates. We thus define the projectors as
{\setlength{\abovedisplayskip}{2pt}%
\setlength{\belowdisplayskip}{2pt}
\begin{gather}
\Pi_e = |\nu_e\rangle\langle\nu_e| \\
\Pi_\mu = |\nu_\mu\rangle\langle\nu_\mu| \\
\Pi_\mathrm{off} = \mathbb{I}-\Pi_e-\Pi_\mu = |\nu_\tau\rangle\langle\nu_\tau|
\end{gather}
where $\Pi_e$ and $\Pi_\mu$ project onto the electron and muon neutrino states, respectively, and $\Pi_\mathrm{off}$ accounts for all outcomes other than $\nu_e$ or $\nu_\mu$. As we are here assuming that the efficiencies in the detection of both electron and muon neutrinos are equal to one, the last operator does in fact correspond to the detection of tau neutrinos $\nu_\tau$. Although $\nu_\tau$ cannot be always directly detected, its information is still contained in the electron and muon neutrino detections: given the number of muonic and electronic neutrinos detected, $M_\mu$ and $M_e$, the number of tauonic neutrinos can in fact be inferred as $M_\tau = M - (M_\nu+M_e)$; as we mentioned before, the total number of neutrinos $M$ arriving at the far-detector of the experiment can be indeed efficiently estimated from the number of events obtained at the near-detector \cite{PhysRevD.87.012001,Alekou:2022}.
An analogous construction applies for a $\overline{\nu}_\mu$ beam, using projectors onto antineutrino flavor eigenstates.
\begin{figure}[h] %h per here se no la mette dove vuole lui
  \centering
    \subfigure{
    \includegraphics[width=0.45
\textwidth, height=0.6\textheight, keepaspectratio]{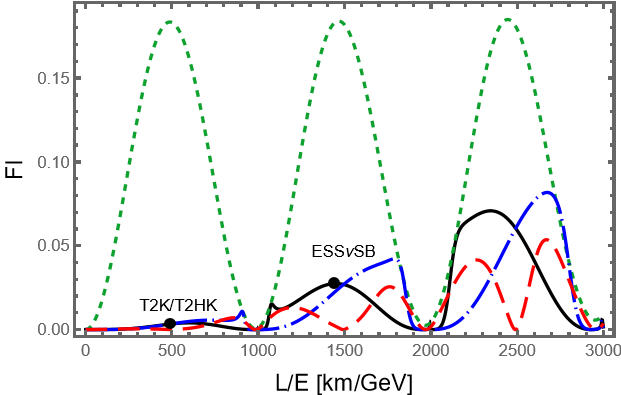}}
      \caption{Classical effective FI corresponding to flavor detection $F_{\nu,\overline{\nu}}(\delta_{CP})$ as a function of the effective baseline $L/E$ and for different values of $\delta_{CP}$: black solid line, $\delta_{CP}=\delta_{CP}^{(\mathrm{NO})}$; red long-dashed line, $\delta_{CP}=90^\circ$; blue dash-dotted line , $\delta_{CP}=0^\circ$. The short-dashed green line corresponds to the effective QFI $H_{\nu,\overline{\nu}}(\delta_{CP})$, while the two black points correspond to the value of the FI for the  experiments T2K/T2HK and ESS$\nu$SB. These have been identified by considering the peak energy $E$ of the neutrino spectrum.}
   \label{FIdeltaCP}
\end{figure}

As before, by exploiting the formula in Eq.~\eqref{eq:fi_app_def} we can evaluate the FIs corresponding to both neutrino experiments, $F_\nu(\delta_{CP})$, and antineutrino experiments, $F_{\bar\nu}(\delta_{CP})$. In this case one can verify that the corresponding values may have a non negligible difference in some regimes, in particular for values of $\delta_{CP}$ corresponding to maximum CP-violation. However it is still true that the two quantities are approximately equal for values of $L/E$ corresponding to probability oscillation maxima, that is in correspondence of the current and future neutrino experiments. We can thus define an effective FI that takes into account both neutrinos and antineutrinos events as $F_{\nu,\overline{\nu}}(\delta_{CP}) =( F_{\nu}(\delta_{CP})+F_{\bar\nu}(\delta_{CP}))/2$.

%Also in this case one can verify how the two quantities are approximately equal in the parameters regime of interest \michela{Ho controllato e in realtà non è vero dipende dal valore di $\delta_{CP}$ per la FI}, and we can thus define an effective FI that takes into account both experiments as $F_{\nu,\overline{\nu}}(\delta_{CP}) =( F_{\nu}(\delta_{CP})+F_{\bar\nu}(\delta_{CP}))/2$.

In Fig.~\ref{FIdeltaCP}, this quantity is seen to depend on both $\delta_{CP}$ and on the location of the maxima, in contrast to the QFI.
As required by Eq.~\eqref{eq:qcrb}, the FI never exceeds the QFI: for all $\delta_{CP}$ values, $F_{\nu,\overline{\nu}}(\delta_{CP})$ remains below the peaks of $H_{\nu,\overline{\nu}}(\delta_{CP})$, indicating that flavor measurements are suboptimal for estimating $\delta_{CP}$ and that the achievable precision is therefore limited by $F_{\nu,\overline{\nu}}(\delta_{CP})$.
The strategy centered on the first oscillation maximum is strongly suboptimal, yielding FI values several orders of magnitude smaller than the quantum Fisher limit, whereas at the second maximum the reduction is only by a factor of about five. Indeed, $F_{\nu,\overline{\nu}}(\delta_{CP})$ typically increases with oscillation order, larger at the second peak than at the first, but smaller than at the third.
In the case of maximal CP violation, the FI departs from the oscillation probability, exhibiting oscillations at approximately twice the frequency. It therefore vanishes at the second oscillation peak and reaches its maximum between probability minima and maxima, while still attaining larger overall values of $F_{\nu,\overline{\nu}}(\delta_{CP})$ for peaks corresponding to higher $L/E$. Apart from this special case, the general trend indicates that experiments targeting the second oscillation maximum, such as ESS$\nu$SB, are inherently better suited for $\delta_{CP}$ determination than first-peak setups like T2K or T2HK, unless the latter achieve substantially higher statistics. This observation confirms earlier analyses \cite{Coloma_2012}, which noted that the second oscillation peak amplifies the relative dependence of the appearance probability on $\delta_{CP}$, further justifying them under the rigorous framework of QET. 

\textit{\textbf{Conclusions}}
We applied QET to study the estimation of $\delta_{CP}$ via neutrino oscillations experiments, computing both the corresponding QFI and classical FI, in order to assess whether the current experimental uncertainty is limited by the information encoded in the neutrino quantum state itself or by the restriction to flavor measurements, which constitute the only feasible detection strategy.
The QFI is insensitive to $\delta_{CP}$ and follows the oscillation pattern of the appearance probability. The corresponding Cramér–Rao bound yields $\sigma \sim 3^\circ$, roughly an order of magnitude smaller than the current best-fit value \cite{Esteban2024NuFit}, indicating that present uncertainties are not limited by the information encoded in the neutrino quantum state.
Instead, we find that the restriction to flavor measurements constitutes a limiting factor for the achievable precision.
The FI varies with $\delta_{CP}$ and grows at higher-order oscillation maxima, revealing enhanced sensitivity at the second and third peaks. At the first maximum, where current experiments operate, the FI is orders of magnitude below the QFI, underscoring the suboptimal nature of flavor measurement strategies.
The planned ESS$\nu$SB experiment, targeting the second peak, achieves a FI approaching the QFI, demonstrating near-optimal extraction of the available quantum information.
Although the largest limitation to the precision originates from the measurement strategy rather than the quantum state itself, it should still be regarded as an intrinsic impediment at the first oscillation peak, unless a substantial increase in statistics becomes feasible.
Building on the present results, a upcoming paper will compute the QFI and FI for additional neutrino oscillation parameters, providing a unified framework to compare their respective sensitivities~\cite{upcoming}.
This work, part of the growing effort to connect high-energy physics with quantum information theory and quantum sensing~\cite{chou2023quantumsensorshighenergy,Beadle:2933206,Afik_2025}, provides a quantitative foundation for optimizing future neutrino oscillation experiments. Our analysis shows that achieving precise $\delta_{CP}$ measurements does not require fundamentally new quantum states, but rather a strategic experimental configuration, specifically, targeting higher oscillation maxima, that maximizes the extraction of existing information.  Furthermore our approach paves the way to future studies, aimed to proper analyze the effect of all possible experimental inefficiencies that we have overlooked in the present analysis. More in general, QET thus emerges as a powerful tool for benchmarking and guiding experimental design in high energy physics, shifting the focus from sheer statistics to strategic information use.

{\em Acknowledgments} - The authors acknowledge insightful discussions with D. Alves, P. Coloma and A. Serafini, and in particular thank F. Terranova for his valuable help and stimulating discussions.

\bibliographystyle{apsrev4-2}
\bibliography{bibliografia}

%
%\appendix
%
\onecolumngrid
\newpage
\begin{center}
 \textbf{\large End Matter}
\end{center}
\twocolumngrid

\appendix

\section{Neutrino oscillations}

Neutrino oscillations in vacuum are a quantum mechanical phenomenon made possible by the existence of non-degenerate neutrino masses ($m_1,m_2,m_3$) and lepton flavor mixing.
The origin of flavor mixing in the lepton sector lies in the mismatch between the orthonormal basis of interacting flavor eigenstates  ($|\nu_e\rangle,|\nu_\mu\rangle,|\nu_\tau\rangle$) and the orthonormal basis of mass eigenstates ($|\nu_1\rangle,|\nu_2\rangle,|\nu_3\rangle$).\par The unitary transformation relating the left-handed flavor eigentstate neutrino fields to the left-handed mass eigenstate neutrino fields is the Pontecorvo-Maki-Nakagawa-Sakata (PMNS) matrix:

\begin{equation}
    \begin{pmatrix}
        \nu_e(x) \\ \nu_\mu(x) \\ \nu_\tau(x)
    \end{pmatrix}_L = \begin{pmatrix}
    U_{e1} \ U_{e2} \ \ U_{e3} \\
    U_{\mu1} \ U_{\mu2} \ \ U_{\mu3} \\
    U_{\tau1} \ U_{\tau2} \ \ U_{\tau3}
    \end{pmatrix} \ \begin{pmatrix}
        \nu_1(x) \\ \nu_2(x) \\ \nu_3(x)
    \end{pmatrix}_L
    \label{eq:mixing}
\end{equation}

Considering neutrinos as Dirac fermions, the PMNS matrix can be parametrized by three mixing angles ($\theta_{12},\theta_{13},\theta_{23}$) and one CP-violating phase $\delta_{CP}$. The explicit form of the PMNS matrix is:
\begin{widetext}
    \begin{equation}
        \resizebox{\textwidth}{!}{$
U=\begin{pmatrix}
 \cos{\theta_{12}} \cos{\theta_{13}} & \sin{\theta_{12}} \cos{\theta_{13}} & e^{-i \delta_{CP} } \sin{\theta_{13}} \\
 -\sin{\theta_{12}}\cos{\theta_{23}}-e^{i \delta_{CP}} \cos{\theta_{12}} \sin{\theta_{13}} \sin{\theta_{23}} & \cos{\theta_{12}} \cos{\theta_{23}}-e^{i \delta_{CP}} \sin{\theta_{12}} \sin{\theta _{13}} \sin{\theta_{23}} & \cos{\theta_{13}} \sin{\theta_{23}} \\
 \sin{\theta_{12}} \sin{\theta_{23}}-e^{i \delta_{CP}} \cos{\theta_{12}} \sin {\theta_{13}} \cos{\theta_{23}} & -\cos{\theta_{12}} \sin{\theta_{23}}-e^{i \delta_{CP}} \sin{\theta_{12}} \sin{\theta_{13}} \cos{\theta_{23}} & \cos{\theta_{13}} \cos{\theta_{23}} \\
\end{pmatrix}
$}
    \end{equation}
\end{widetext}
In the case in which neutrinos are Majorana fermions two additional phases appear but they do not affect oscillation probabilities.\par
From Eq. \eqref{eq:mixing}, considering the quantum excitations of the fields, i.e. the particles quantum state, we can write:
\begin{equation}
    |\nu_{\alpha}\rangle=\sum_i U_{\alpha i}^* |\nu_i\rangle  \quad \text{with} \quad \alpha=e,\mu,\tau,
    \label{eq: stato 0}
\end{equation}
where $|\nu_i\rangle$ denotes the mass eigenstate with eigenvalue $m_i$.\par 
In vacuum each mass eigenstate $|\nu_i\rangle$ is an eigenstate of the free Hamiltonian with eigenvalue $E_i=\sqrt{p^2+m_i^2}$, where $p$ is the momentum of the neutrino beam and wlog we consider it to be the same for all mass eigenstates:
\begin{equation}
    |\nu_i(t)\rangle=e^{-i E_it}|\nu_i\rangle,
\end{equation}
thus propagation modifies the original coherent superposition in Eq. \eqref{eq: stato 0}, which is no longer a pure flavor eigenstate. Assuming that the neutrino is generated at time $t_0=0$ in a flavor eigenstate $|\nu_{\alpha}\rangle$, the neutrino state at a generic time $t$ can be written as:
\begin{equation}
    |\nu_\alpha(t)\rangle=\sum_i U_{\alpha i}^* e^{-i E_it} |\nu_i\rangle=\sum_i U_{\alpha i}^* e^{-i E_it} \sum_\beta U_{\beta i} |\nu_\beta\rangle,
    \label{eq : change basis}
\end{equation}
where we have rewritten the evolved state in the flavor eigenbasis since in this work we are interested in flavor measurements. From this expression we observe that the neutrino state at time $t$ depends on the parameters of the PMNS matrix, as well as on $E_i$ and $t$.\par
The probability that at time $t$ the neutrino of flavor $\alpha$ has oscillated into a neutrino of flavor $\beta$ is given by:
\begin{equation}
\begin{split}
    P(|\nu_\alpha\rangle \to |\nu_\beta\rangle)&= |\langle\nu_\beta|\nu_\alpha(t)\rangle|^2=\left|\sum_i U_{\beta i}U_{\alpha i}^* e^{-iE_it}\right|^2.
    \label{eq:prob-oscil}
    \end{split}
\end{equation}
Since neutrinos are ultra relativistic, the approximation  $E_i\approx p+m_i^2/2E$ with $E\approx p$ is valid and the probability can be rewritten as:
\begin{equation}
\begin{split}
    P(|\nu_\alpha\rangle \to |\nu_\beta\rangle)&=\sum_{i,j} U_{\beta i}U_{\alpha i}^* U_{\beta j}^*U_{\alpha j} e^{i\frac{\Delta m_{ji}^2 L}{2E}},
    \end{split}\label{eq:prob}
\end{equation}
where $\Delta m_{ij}^2 \equiv m_j^2-m_i^2$ and $L\approx ct$. Thus the oscillation probability depends on the PMNS parameters, on the squared mass differences $\Delta m_{ij}^2$ and on the ratio $L/E$ between the distance traveled by the neutrino and its energy. In experiments $L$ is referred to as the baseline, i.e the distance between the production and detection points of the neutrino beam.\par
Expanding Eq. \eqref{eq:prob} we obtain:
\begin{align}
\begin{split}
    P(|\nu_\alpha\rangle \to |\nu_\beta\rangle)&=\delta_{\alpha\beta} \\&-4 \sum_{i<j} \text{Re}[U_{\alpha i}U_{\beta i}^*U_{\alpha j}^*U_{\beta j}]\sin^2\biggl(\frac{\Delta m_{ji}^2 L}{4E}\biggl)\\&+ 2 \sum_{i<j} \text{Im}[U_{\alpha i}U_{\beta i}^*U_{\alpha j}^*U_{\beta j}]\sin\biggl(\frac{\Delta m_{ji}^2 L}{2E}\biggl).
    \end{split}\label{eq:prob2}
\end{align}

We can clearly see that the term "neutrino oscillations" is used since the probability of flavor changing is a sum of sinusoidal and sine squared functions of the variable $L/E$. Moreover we observe that the phase $\delta_{CP}$ enters through the imaginary parts of the PMNS matrix combinations. \par

\par

\end{document}